# A Novel Compensation Algorithm for Thickness Measurement Immune to Lift-Off Variations Using Eddy Current Method

Mingyang Lu, Liyuan Yin, Anthony J. Peyton, and Wuliang Yin, *Senior Member, IEEE*

*Abstract*—Lift-off variation causes errors in the eddy current thickness measurements of metallic plates. In this paper, we have developed an algorithm that can compensate for this variation and produce an index that is linked to the thickness, but is virtually independent of lift-off. This index, termed as the compensated peak frequency, can be obtained from the measured multifrequency inductance spectral data using the algorithm we developed in this paper. This method has been derived through mathematical manipulation and verified by both the simulation and experimental data. Accuracy in the thickness measurements at different lift-offs proved to be within 2%.

## I. INTRODUCTION

THE thickness of metallic plates can be measured using both multifrequency and pulse eddy current testing methods [1]–[11]. However, both the methods suffer from errors caused by the so-called lift-off effect. To address this issue, researchers have investigated a range of methods such as using different signal processing, feature extraction [12], [13], sensor structure [14]–[16], and detection principles [17]–[20]. Multifrequency eddy current sensing in the context of nondestructive testing applications has been the focus of the authors' research in recent years. Conductivity and permeability depth profiling [21], [22] and noncontact microstructure monitoring [23]–[29] have been explored.

In a recent development [30], we proposed a novel design of an eddy current sensor, composed of three coils and operating as an axial gradiometer interrogated with a multifrequency waveform. The difference in the peak frequencies of the impedance/inductance spectra from the gradiometer was used for thickness evaluation and showed to be virtually immune to lift-off variations.

Manuscript received April 12, 2016; revised June 8, 2016; accepted July 16, 2016. Date of publication August 30, 2016; date of current version November 8, 2016. This work was supported by the U.K. Engineering and Physical Sciences Research Council. The Associate Editor coordinating the review process was Dr. Sasan Bakhtiari. (*Corresponding author: Wuliang Yin.*)

M. Lu, A. J. Peyton, and W. Yin are with the School of Electrical and Electronic Engineering, The University of Manchester, Manchester, M60 1QD, U.K. (e-mail: wuliang.yin@manchester.ac.uk).

L. Yin is with Kunming Science and Technology University, Kunming 650051, China.

Color versions of one or more of the figures in this paper are available online at http://ieeexplore.ieee.org.

Digital Object Identifier 10.1109/TIM.2016.2600918

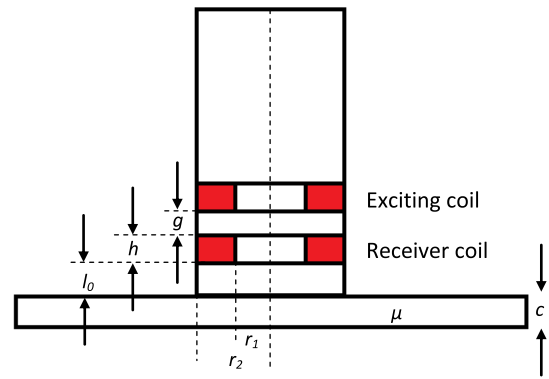

Fig. 1. Sensor configuration.

In this paper, we consider a simpler structure consisting of just one coil pair. This method has the advantages of a less complicated mechanical configuration as well as improved accuracy, and avoids the need for a precise magnetic balance, as in the case of the gradiometer. The new technique exploits two facts. First, the peak frequency of the inductance spectral signal decreases with increased lift-off, and second, the magnitude of the signal decreases with increased lift-off. An algorithm has been proposed to compensate for the change in the peak frequency. Theoretically, simulation and experiments show that the compensated peak frequency is nearly lift-off independent and therefore provide accurate thickness estimation.

## II. SENSOR DESCRIPTION

The sensor is composed of two coils, one as excitation and the other as receiver, both of which have the same dimensions and are arranged coaxially. A schematic plot of the sensor is shown in Fig. 1, with its dimensions in Table I.

The connection of this sensor to an impedance analyzer (SL 1260) is shown in Fig. 2.

## III. THEORETICAL DERIVATION OF THE COMPENSATED PEAK FREQUENCY

Previously, we have proved that the peak frequency decreases with increased lift-off [30]. For completeness, the main steps are summarized in this paper. It is also common knowledge that the signal amplitude also decreases with increased lift-off. Therefore, we hypothesize that an algorithm can be developed to compensate for the variation in the peak frequency with the signal amplitude. In the following, we will derive such an algorithm.

TABLE I
COIL PARAMETERS

| | |
|---|---|
| $r_1$ | 11.8 mm |
| $r_2$ | 12 mm |
| $lo$ (lift-off) | 0.5 mm |
| $h$ (height) | 3 mm |
| $g$ (gap) | 1 mm |
| Number of turns    N1 = N2 | 20 |

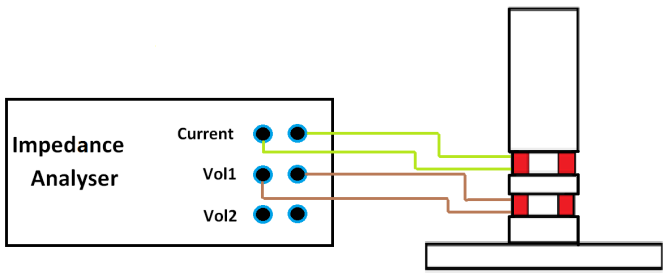

Fig. 2.   Experimental wiring schematic design.

We start with the Dodd and Deeds analytical solution [31], which describes the inductance change in an air-core coil caused by a layer of nonmagnetic, metallic plates. Other similar formulas exist [32]. The difference in the complex inductance is $\Delta L(\omega) = L(\omega) - L_A(\omega)$, where the coil inductance above a plate is $L(\omega)$ and $L_A(\omega)$ is the inductance in free space.

The formulas of Dodd and Deeds [31] are

$$\Delta L(\omega) = K \int_0^\infty \frac{P^2(\alpha)}{\alpha^6} A(\alpha)\phi(\alpha) d\alpha \qquad (1)$$

where

$$\phi(\alpha) = \frac{(\alpha_1 + \alpha)(\alpha_1 - \alpha) - (\alpha_1 + \alpha)(\alpha_1 - \alpha)e^{2\alpha_1 c}}{-(\alpha_1 - \alpha)(\alpha_1 - \alpha) + (\alpha_1 + \alpha)(\alpha_1 + \alpha)e^{2\alpha_1 c}} \qquad (2)$$

$$\alpha_1 = \sqrt{\alpha^2 + j\omega\sigma\mu_0} \qquad (3)$$

$$K = \frac{\pi\mu_0 N^2}{h^2(r_1 - r_2)^2} \qquad (4)$$

$$P(\alpha) = \int_{\alpha r_1}^{\alpha r_2} x J_1(x) dx \qquad (5)$$

$$A(\alpha) = e^{-\alpha(2l_0 + h + g)}(e^{-2\alpha h} + 1) \qquad (6)$$

where $\mu_0$ denotes the permeability of free space; $N$ denotes the number of turns in the coil; $r_1$ and $r_2$ denote the inner and outer radii of the coil; $l_0$ and $h$ denote the lift-off and the height of the coil; and $c$ denotes the thickness of the plate.

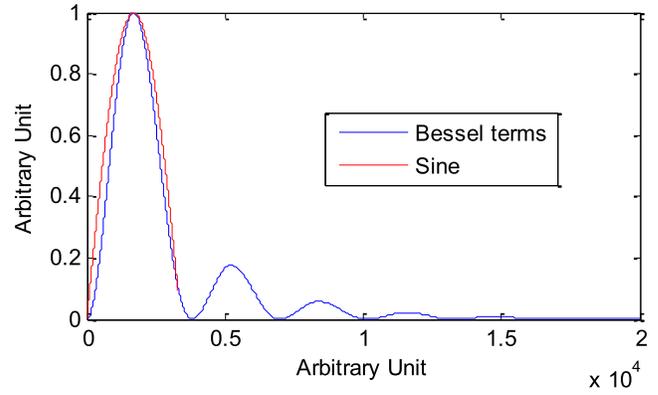

Fig. 3.   Approximation of the Bessel term with a sinusoid.

Equations (1)–(6) can be approximated based on the fact that $\phi(\alpha)$ varies slowly with $\alpha$ compared with the rest of the integrand, which reaches its maximum at a characteristic spatial frequency $\alpha_0$ (Fig. 3). The approximation is to evaluate $\phi(\alpha)$ at $\alpha_0$ and take it outside of the integral

$$\Delta L(\omega) = \phi(\alpha_0)\Delta L_0 \qquad (7)$$

where

$$\phi(\alpha_0) = \frac{(\alpha_1 + \mu\alpha_0)(\alpha_1 - \mu\alpha_0) - (\alpha_1 + \mu\alpha_0)(\alpha_1 - \mu\alpha_0)e^{2\alpha_1 c}}{-(\alpha_1 - \mu\alpha_0)(\alpha_1 - \mu\alpha_0) + (\alpha_1 + \mu\alpha_0)(\alpha_1 + \mu\alpha_0)e^{2\alpha_1 c}} \qquad (8)$$

$$\Delta L_0 = K \int \frac{P^2(\alpha)}{\alpha^6} A(\alpha) d\alpha. \qquad (9)$$

Note that in (7), the sensor phase signature is solely determined by $\phi(\alpha_0)$, which includes conductivity, the thickness of the conducting plate, and $\alpha_0$. $\Delta L_0$ is the overall magnitude of the signal, which is strongly dependent on the coil geometrical parameters but independent of the thickness and electromagnetic properties of the plate.

Substituting $e^{2\alpha_1 c}$ with $1 + 2\alpha_1 c$, and considering (3), (8) becomes

$$\phi(\alpha_0) \approx \frac{j\omega\sigma\mu_0 c}{j\omega\sigma\mu_0 c + 2\alpha_0^2 c + 2\alpha_0 + 2\alpha_0\alpha_1 c}. \qquad (10)$$

Assigning

$$\omega_1 = \frac{2\alpha_0^2 c + 2\alpha_0}{\sigma\mu_0 c}. \qquad (11)$$

Equation (10) can be expressed as

$$\phi(\alpha) = \frac{j\omega/\omega_1}{j\omega/\omega_1 + 1 + 2\alpha_0\alpha_1 c/(2\alpha_0^2 c + 2\alpha_0)}. \qquad (12)$$

In (12), it can be seen that the peak frequency for the first-order system is approximately $\omega_1$ and from (11) it is concluded that the peak frequency increases with $\alpha_0$.

Suppose a lift-off variation of $l_0$ is introduced, from (6), we can see that an increase of $l_0$ in lift-off is equivalent to multiplying a factor $e^{-2\alpha l_0}$

$$A(\alpha) = e^{-\alpha(2l_0 + h + g)}(e^{-2\alpha h} + 1). \qquad (13)$$

Due to the fact that $\Delta L_0 = K \int ((P^2(\alpha))/(\alpha^6))A(\alpha)d\alpha$ peaks at $\alpha_0$ and that the squared Bessel term $P^2(\alpha)$ is the main contributor, a simple function $\sin^2((\alpha\pi)/(2\alpha_0))$ with its maximum at $\alpha_0$ is used to approximate $\Delta L_0$

$$\mathrm{Im}(\Delta L_0) \approx \mathrm{Im}(\Delta L_m)e^{-2\alpha l_0}\sin^2\left(\frac{\alpha\pi}{2\alpha_0}\right) \qquad (14)$$

where $\Delta L_m$ denotes the magnitude of the inductance change when the lift-off is zero.

This simplification is applied to obtain an analytical solution for $\alpha_0$.

The shift in $\alpha_0$ due to the effect of lift-off can be predicted as follows.

The new $\alpha$ should maximize $e^{-2\alpha l_0}\sin^2(\alpha\pi/2\alpha_0)$ and therefore $e^{-\alpha l_0}\sin(\alpha\pi/2\alpha_0)$.

The maximum can be obtained by finding the stationary point for $e^{-\alpha l_0}\sin(\alpha\pi/2\alpha_0)$.

Let

$$\left(e^{-\alpha l_0}\sin\left(\frac{\alpha\pi}{2\alpha_0}\right)\right)' = -l_0 \cdot e^{-\alpha l_0}\sin\left(\frac{\alpha\pi}{2\alpha_0}\right)$$
$$+ \frac{\pi}{2\alpha_0}e^{-\alpha l_0}\cos\left(\frac{\alpha\pi}{2\alpha_0}\right) = 0.$$

And through some mathematical manipulations, a new equation can be obtained

$$\frac{\alpha\pi}{2\alpha_0} = \tan^{-1}\left(\frac{\pi}{2\alpha_0 l_0}\right).$$

With a small lift-off variation, $\alpha_0 l_0 \ll 1$ holds and the right side can be approximated as $(\pi/2) - ((2\alpha_0 l_0)/\pi)$.

Therefore, the revised $\alpha_0$, $\alpha_{0r}$ is

$$\alpha_{0r} = \alpha_0 - \frac{4\alpha_0^2 l_0}{\pi^2}. \qquad (15)$$

Combining (11) with (15), $\omega_1$ becomes

$$\omega_1 \approx \frac{2(\alpha_0^2\pi^4 - 8\pi^2\alpha_0^3 l_0 + 16\alpha_0^4 l_0^2)c + 2(\alpha_0\pi^4 - \pi^2 4\alpha_0^2 l_0)}{\pi^4\sigma\mu_0 c}. \qquad (16)$$

Combining (14) with (15), $\mathrm{Im}(\Delta L_0)$ becomes

$$\mathrm{Im}(\Delta L_0) = \mathrm{Im}(\Delta L_m)e^{-2\left(\alpha_0 - \frac{4\alpha_0^2 l_0}{\pi^2}\right)l_0}\cos^2\left(\frac{2\alpha_0 l_0}{\pi}\right)$$
$$= \mathrm{Im}(\Delta L_m)e^{-2\left(\alpha_0 - \frac{4\alpha_0^2 l_0}{\pi^2}\right)l_0}\left(\frac{\cos\left(\frac{4\alpha_0 l_0}{\pi}\right) + 1}{2}\right).$$

Considering $\alpha_0 l_0 \ll 1$ and based on small-angle approximation $\cos(\theta) \approx 1 - ((\theta^2)/2)$, $\cos((4\alpha_0 l_0)/\pi)$ is substituted with $1 - (((((4\alpha_0 l_0)/\pi))^2)/2)$.

$\mathrm{Im}(\Delta L_0)$ becomes

$$\mathrm{Im}(\Delta L_0) = \mathrm{Im}(\Delta L_m)e^{-2\left(\alpha_0 - \frac{4\alpha_0^2 l_0}{\pi^2}\right)l_0}\left(1 - \frac{4\alpha_0^2 l^2}{\pi^2}\right).$$

Substituting $(1 - ((4\alpha_0^2 l^2)/\pi^2))$ with $e^{-((4\alpha_0^2 l^2)/\pi^2)}$

$$\mathrm{Im}(\Delta L_0) = \mathrm{Im}(\Delta L_m)e^{-2\left(\alpha_0 - \frac{4\alpha_0^2 l_0}{\pi^2}\right)l_0}e^{-\frac{4\alpha_0^2 l^2}{\pi^2}}$$
$$= \mathrm{Im}(\Delta L_m)e^{-2\left(\alpha_0 - \frac{2\alpha_0^2 l_0}{\pi^2}\right)l_0}. \qquad (17)$$

Taking the natural logarithmic operation of both the sides, we arrive at

$$\ln\frac{\mathrm{Im}(\Delta L_0)}{\mathrm{Im}(\Delta L_m)} = -2\left(\alpha_0 - \frac{2\alpha_0^2 l_0}{\pi^2}\right)l_0. \qquad (18)$$

And further

$$4\alpha_0^2 l_0^2 - 2\pi^2\alpha_0 l_0 - \pi^2\ln\frac{\mathrm{Im}(\Delta L_0)}{\mathrm{Im}(\Delta L_m)} = 0.$$

This is now a quadratic equation with $\alpha_0 l_0$ as its variable. Therefore, the solution for $\alpha_0 l_0$ is

$$\alpha_0 l_0 = \frac{\pi^2 - \sqrt{\pi^4 + 4\pi^2\ln\frac{\mathrm{Im}(\Delta L_0)}{\mathrm{Im}(\Delta L_m)}}}{4}. \qquad (19)$$

The other solution

$$\alpha_0 l_0 = \frac{\pi^2 + \sqrt{\pi^4 + 4\pi^2\ln\frac{\mathrm{Im}(\Delta L_0)}{\mathrm{Im}(\Delta L_m)}}}{4}$$

does not satisfy the small lift-off condition $\alpha_0 l_0 \ll 1$ and therefore is discarded.

From (19), lift-off can be estimated as

$$l_0 = \frac{\pi^2 - \sqrt{\pi^4 + 4\pi^2\ln\frac{\mathrm{Im}(\Delta L_0)}{\mathrm{Im}(\Delta L_m)}}}{4\alpha_0}. \qquad (20)$$

Combining (16) with (20), the peak frequency with a lift-off of $l_0$ becomes

$$\omega_1 = \frac{2\alpha_0^2 C\left(\pi^2 + 4\ln\frac{\mathrm{Im}(\Delta L_0)}{\mathrm{Im}(\Delta L_m)}\right) + 2\alpha_0\pi\sqrt{\pi^2 + 4\ln\frac{\mathrm{Im}(\Delta L_0)}{\mathrm{Im}(\Delta L_m)}}}{\pi^2\sigma\mu_0 c}. \qquad (21)$$

Equation (21) becomes a quadratic equation with an unknown $\alpha_0$

$$2c\left(\pi^2 + 4\ln\frac{\mathrm{Im}(\Delta L_0)}{\mathrm{Im}(\Delta L_m)}\right)\alpha_0^2$$
$$+ 2\pi\sqrt{\pi^2 + 4\ln\frac{\mathrm{Im}(\Delta L_0)}{\mathrm{Im}(\Delta L_m)}}\alpha_0 - \omega_1\pi^2\sigma\mu_0 c = 0.$$

And the solution is (22), as shown at the bottom of this page.

$$\alpha_0 = \frac{-\pi\sqrt{\pi^2 + 4\ln\frac{\mathrm{Im}(\Delta L_0)}{\mathrm{Im}(\Delta L_m)}} + \sqrt{\pi^2(2\omega_1\sigma\mu_0 c^2 + 1)\left(\pi^2 + 4\ln\frac{\mathrm{Im}(\Delta L_0)}{\mathrm{Im}(\Delta L_m)}\right)}}{2c\left(\pi^2 + 4\ln\frac{\mathrm{Im}(\Delta L_0)}{\mathrm{Im}(\Delta L_m)}\right)}$$
$$= \frac{\pi(\sqrt{(1 + 2\omega_1\sigma\mu_0 c^2)} - 1)}{2c\sqrt{\pi^2 + 4\ln\frac{\mathrm{Im}(\Delta L_0)}{\mathrm{Im}(\Delta L_m)}}} \qquad (22)$$

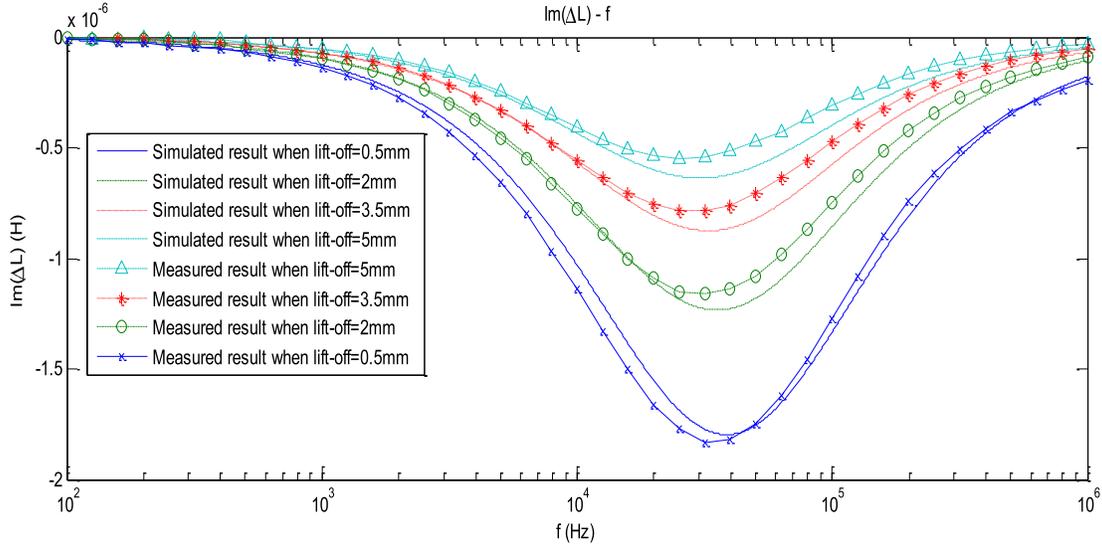

Fig. 4. Simulation and experimental results of imaginary parts of $\Delta L$ for a 22-$\mu$m aluminum plate at a range of lift-offs.

Therefore, by combining (11) with (22), the original peak frequency (peak frequency prior to introducing the lift-off $l_0$) can be obtained as in (23), shown at the bottom of this page.

It can be seen in (23) that through a compensation scheme, and using the knowledge of the peak frequency and the amplitude at a certain lift-off, the original peak frequency (peak frequency prior to introducing the lift-off $l_0$) can be recovered.

Further approximation can be carried out by considering $\alpha_0 c \ll 1$. The peak frequency of the imaginary part of the inductance in (21) becomes

$$\omega_1 \approx \frac{2\alpha_0 \sqrt{\pi^2 + 4\ln\frac{\mathrm{Im}(\Delta L_0)}{\mathrm{Im}(\Delta L_m)}}}{\pi \sigma \mu_0 c}. \tag{24}$$

Equation (22) becomes

$$\alpha_0 \approx \frac{\pi \sigma \mu_0 \omega_1 c}{2\sqrt{\pi^2 + 4\ln\frac{\mathrm{Im}(\Delta L_0)}{\mathrm{Im}(\Delta L_m)}}}. \tag{25}$$

Therefore, the compensated peak frequency (peak frequency prior to introducing the lift-off $l_0$) becomes

$$\omega_0' \approx \frac{2\alpha_0}{\sigma \mu_0 c} = \frac{\pi \omega_1}{\sqrt{\pi^2 + 4\ln\frac{\mathrm{Im}(\Delta L_0)}{\mathrm{Im}(\Delta L_m)}}}. \tag{26}$$

So, the thickness reduces to

$$c = \frac{2\alpha_0}{\sigma \mu_0 \omega_0'} = \frac{2\alpha_0 \sqrt{\pi^2 + 4\ln\frac{\mathrm{Im}(\Delta L_0)}{\mathrm{Im}(\Delta L_m)}}}{\pi \sigma \mu_0 \omega_1}. \tag{27}$$

Therefore from (27) it can be seen that as lift-off increases, the measured peak frequency decreases, but the numerator term also decreases to compensate for this, so that the compensated peak frequency and accurate thickness can be recovered.

## IV. SIMULATIONS AND EXPERIMENTS

Experiments and simulations were carried out to verify the performance of the compensation algorithm; the compensated peak frequency and the compensated thickness measurements at different lift-offs were compared. Here, the imaginary part of the inductance is defined from the mutual impedance of the coils

$$\begin{aligned}
\mathrm{Im}(\Delta L) &= \mathrm{Im}\left(\frac{Z(f) - Z_{\mathrm{air}}(f)}{j2\pi f}\right) \\
&= \mathrm{Re}\left(\frac{-(Z(f) - Z_{\mathrm{air}}(f))}{2\pi f}\right)
\end{aligned} \tag{28}$$

where $Z(f)$ denotes the impedance of the coil in the presence of a metallic plate, while $Z_{\mathrm{air}}(f)$ is that of the coil in air.

### A. Simulations

The simulated sensor configuration is shown in Table I. The simulated targets are aluminum plates with a conductivity of 38.2 MS/m and thickness of 22 and 44 $\mu$m under varying lift-offs 0.5, 2, 3.5, and 5 mm. The simulations were realized by a custom developed solver using MATLAB. The solver can be used to calculate the Dodd and Deeds solution [31] (1)–(6)

$$\begin{aligned}
\omega_0 &= \frac{2\alpha_0^2 c + 2\alpha_0}{\sigma \mu_0 c} \\
&= \frac{\pi^2 (\sqrt{(1 + 2\omega_1 \sigma \mu_0 c^2)} - 1)^2 + 2\pi (\sqrt{(1 + 2\omega_1 \sigma \mu_0 c^2)} - 1)\sqrt{\pi^2 + 4\ln\frac{\mathrm{Im}(\Delta L_0)}{\mathrm{Im}(\Delta L_m)}}}{2\sigma \mu_0 c^2 \left(\pi^2 + 4\ln\frac{\mathrm{Im}(\Delta L_0)}{\mathrm{Im}(\Delta L_m)}\right)}
\end{aligned} \tag{23}$$

TABLE II
THICKNESS MEASUREMENTS FOR DIFFERENT THICKNESSES AND LIFT-OFFS

| Lift-off (mm) | Actual thickness (µm) | Thickness calculated from original as measured peak frequency (µm) | Thickness calculated from Compensated peak frequency by the simple 2 coils probe (µm) | Thickness tested from previous Triple-Coil probe (µm) |
|---|---|---|---|---|
| 1.5 | 22 | 23.1 | 22.2 | 22.14 |
| 1.5 | 44 | 46.3 | 44.3 | 43.21 |
| 2 | 22 | 23.5 | 22.3 | 21.42 |
| 2 | 44 | 47 | 44.2 | 44.83 |
| 3.5 | 22 | 23.8 | 22.2 | 21.22 |
| 3.5 | 44 | 47.9 | 44.4 | 44.95 |

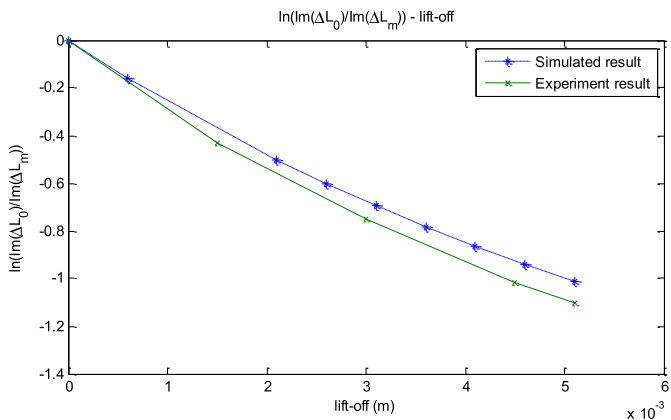

Fig. 5. Simulation and experimental results. Natural logarithmic of $\text{Im}(\Delta L)$ ratio for the 22-$\mu$m sample at a range of lift-offs.

and calculate the thickness and peak frequency using (24)–(27). The solver can take a range of different parameters such as frequency, sample conductivity, thickness, and lift-off. In addition, the solver have been converted and packaged into an executable program.

### B. Experimental Setup

The sensor configuration and the test pieces are the same as that of the simulations. And the multifrequency response of the sensor was obtained by an SL 1260 impedance analyzer with frequency sweeping mode.

It can be seen from Fig. 4 that the peak frequency decreases as lift-off increases and, at the same time, the magnitude of the signal decreases with increased lift-offs. Fig. 5 shows the experimental result.

As can be seen from Fig. 6, the compensated peak frequency decreases following initial lift-off but remains almost constant for larger lift-offs, i.e., peak frequency is virtually immune to lift-off variations for larger lift-offs. Tests were carried out to verify this method and the thickness measurement results are shown in Table II for varying thicknesses and lift-offs. Here, the small thickness is defined as $c \ll 1/\alpha_0$, so it depends on the size of the coil. In our case, this value is generally <1 mm. It can be seen from Fig. 6 that both the compensated and the uncompensated peak frequency are smaller than the actual peak frequency. Since the thickness is inversely proportional to the peak frequency [see (27)], therefore, the calculated thickness is larger than the actual thickness.

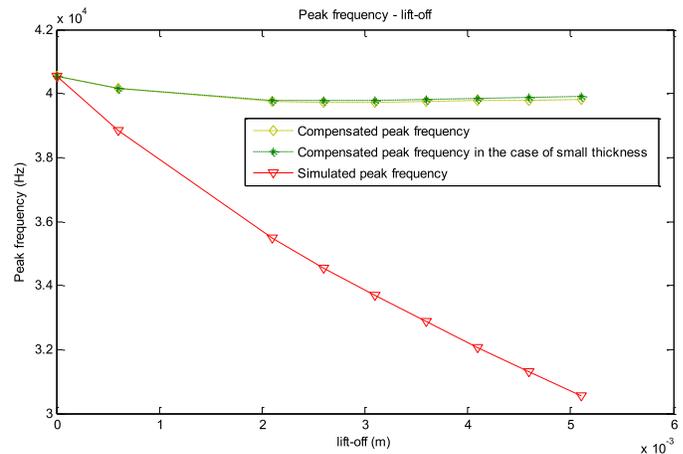

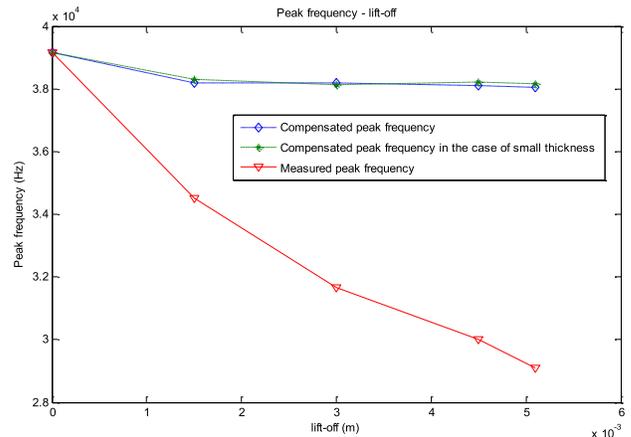

Fig. 6. Comparisons of as-measured (uncompensated) and compensated peak frequencies for the 22-$\mu$m aluminum plate at a range of lift-offs. (a) Simulation results. (b) Experimental results.

In addition, a comparison of the thickness tested from the previous Triple-Coil probe is also added in Table II. It can be seen that the method reported in this paper has in general improved accuracy in thickness measurement and the coil has a simpler structure.

## V. Conclusion

This paper has considered a compensation scheme for reducing the errors in multifrequency eddy current thickness measurements of metallic plates. The peak frequency decreases as lift-offs increase and an algorithm has been developed that can compensate for this variation frequency and produce an index that is linked to thickness but is virtually independent of lift-off. Both the simulation and experimental results show that the compensated peak frequency is almost immune to the lift-off variations. This is an important feature as a lift-off variation is unavoidable in many practical applications. Although the algorithm involved is slightly more complicated, this new approach has the advantages of a less complicated mechanical configuration as well as improved accuracy, and avoids the need for a precise magnetic balance, as in the case of the three-coil configuration [30].

An SL 1260 impedance analyzer working in a swept frequency mode was used to acquire the multifrequency data in this paper. However, multifrequency impedances can also be abstracted simultaneously using composite multisine waveform excitation followed by fast Fourier transform operations as in [33].

**Mingyang Lu** is currently pursuing the Ph.D. degree under the supervision of Dr. W. Yin with the School of Electrical and Electronic Engineering, University of Manchester, Manchester, U.K.

He is involved in developing a finite-element model to solve electromagnetic simulation by taking into account the random geometry and material properties (including microstructure). He is also developing software to increase efficiency of simulations to avoid remeshing, for example, to consider a moving sensor as a moving effective field above a flaw (nondestructive testing application).

**Liyuan Yin**, photograph and biography not available at the time of publication.

**Anthony J. Peyton** received the B.Sc. degree in electrical and electronics engineering and the Ph.D. degree from the University of Manchester Institute of Science and Technology (UMIST), Manchester, U.K., in 1983 and 1986, respectively.

He was a Principal Engineer with Kratos Analytical Ltd., Manchester, in 1989, where he was involved in developing precision electronic instrumentation systems for magnetic sector and quadrupole mass spectrometers. He joined the Process Tomography Group, UMIST, where he was a Lecturer. In 1996, he was a Senior Lecturer with Lancaster University, Lancaster, U.K., where he was a Reader in Electronic Instrumentation in 2001, and a Professor in 2004. Since 2004, he has been a Professor of Electromagnetic Tomography Engineering with the University of Manchester, Manchester. His current research interests include instrumentation, applied sensor systems, and electromagnetics.

**Wuliang Yin** (M'05–SM'06) was appointed as an MT Sponsored Lecturer with the School of Electrical and Electronic Engineering, University of Manchester, Manchester, U.K., in 2012, and was promoted to Senior Lecturer in 2016. He has authored one book, more than 140 papers, and was granted more than ten patents in the area of electromagnetic (EM) sensing and imaging. He currently leads several major grants from U.K. government bodies including TSB and EPSRC and is involved in several European Union projects.

Dr. Yin was a recipient of the 2014 and 2015 Williams Award from the Institute of Materials, Minerals and Mining for his contribution in applying EM imaging in the steel industry. He also received the Science and Technology Award from the Chinese Ministry of Education in 2000.